\documentclass[12pt]{article}
\usepackage{amsmath,amsfonts}
\usepackage{graphicx}
\input epsf

\makeatletter\@addtoreset{equation}{section}\makeatother

\def\be{\begin{equation}}
\def\ee{\end{equation}}
\def\bea{\begin{eqnarray}}
\def\eea{\end{eqnarray}}

\makeatletter\@addtoreset{equation}{section}\makeatother

\renewcommand{\title}[1]{\vbox{\center\LARGE{#1}}\vspace{5mm}}
\renewcommand{\author}[1]{\vbox{\center#1}\vspace{5mm}}
\newcommand{\address}[1]{\vbox{\center\em#1}}
\newcommand{\email}[1]{\vbox{\center\tt#1}\vspace{5mm}}

\begin{document}

\unitlength = .8mm

\begin{titlepage}
\begin{center}
\hfill \\
\hfill \\
\vskip 1cm

\title{ZZ boundary states and fragmented $AdS_2$ spaces}

\author{Simone Giombi$^{a}$ and Xi Yin$^{b}$}

\address{Center for the Fundamental Laws of Nature
\\Jefferson Physical Laboratory, Harvard University,\\
Cambridge, MA 02138 USA}

\email{$^a$giombi@physics.harvard.edu,
$^b$xiyin@fas.harvard.edu}

\end{center}

\abstract{
In this paper we show that Liouville gravity on the strip with Zamolodchikov-Zamolodchikov (ZZ) boundary conditions has a semi-classical interpretation in terms of fragmented $AdS_2$ spacetime geometries. Further, we study the three-point functions of the ZZ boundary primaries, and show that they are dominated by multi-$AdS_2$ instantons in the classical limit.}

\vfill

\end{titlepage}

\eject

\tableofcontents

\section{Introduction}

Two-dimensional gravity has been extensively explored in the past 30 years, both as the worldsheet description of
string theories and as a toy model for higher dimensional quantum gravity (for a review see \cite{Ginsparg:1993is}
and references therein). Quantum gravity
in $AdS_2$, which is expected to be related to extremal black holes, remains mysterious
\cite{Strominger:1998yg,Maldacena:1998uz,Strominger:2003tm,Hartman:2008dq,Sen:2008vm,Sen:2008yk}. In this paper,
we shall take the viewpoint that ``pure" quantum gravity in $AdS_2$ is described by Liouville gravity, with Zamolodchikov-Zamolodchikov (ZZ) boundary conditions \cite{Zamolodchikov:2001ah}.\footnote{For reviews of Liouville theory see for example \cite{Seiberg:1990eb} \cite{Nakayama:2004vk}.} The states of quantum gravity in global $AdS_2$ will be the boundary primaries of ZZ.
This proposal will be validated by finding the semi-classical interpretation of these states and their
correlation functions. We will see that in the semi-classical limit, the ZZ boundary primaries describe
``fragmented" $AdS_2$'s, i.e. several global $AdS_2$'s ``attached" along their boundaries.
The correlation functions of the ZZ boundary primaries will be dominated by the contribution from
classical instantons, which are several Poincar\'e discs suitably ``glued" together along parts of their boundaries.

We analyze the quantum corrections to the two-fragmented $AdS_2$ using the exact bulk-boundary three point functions
on the disc. The radius of the $AdS_2$ solution is large in the semi-classical (weak coupling) limit of Liouville gravity. In this limit, the Liouville theory has either large positive central charge $c_L$ (with real background
charge $Q$), or large negative $c_L$ (with imaginary $Q$). In the $c_L>0$ case, we find that quantum corrections erase one of the
two $AdS_2$'s. In the $c_L<0$ case, the two-fragmented $AdS_2$ survive in the quantum theory.

From the point of view of $AdS_2/CFT_1$ correspondence, our results suggest that the ``$CFT_1$''
dual to pure Liouville gravity in $AdS_2$ comprises a single copy of Virasoro alegbra and a finite set of primary states
-- those of Liouville theory on a strip with ZZ boundary conditions. The correlation functions of these
primaries can in principle be computed exactly using bootstrap methods, which then completely characterizes
the theory.

The paper is organized as follows. In section 2 we first review the ZZ boundary conditions and boundary primaries.
We will then probe the ``geometry" of the semi-classical limit of the ZZ boundary primary using a bulk primary
operator, and show that the boundary primaries can be identified as fragmented $AdS_2$'s. In section 3
we study the three-point functions of the boundary primaries. Once again using the bulk primary probe, we will find that
in the semi-classical limit the bulk-boundary four-point function is dominated by an instanton solution
interpolating fragmented $AdS_2$'s.

\section{ZZ boundary primaries as fragmented $AdS_2$}

\subsection{ZZ boundary conditions and boundary primaries in Liouville theory}

We work in the convention of \cite{Zamolodchikov:2001ah}, and write the Liouville Lagrangian density
(in a flat background metric) as
\begin{equation}
{\cal L} = \frac{1}{4\pi}(\partial_a \phi)^2 + \mu e^{2b\phi}.
\label{Liouville-action}
\end{equation}
The background charge is $Q=b+1/b$, and the central charge of the Liouville CFT is
given by $c=1+6Q^2$. Depending on whether $b$ is real or purely imaginary, the central charge
$c$ is greater or less than 1. If $b$ is imaginary, we may retain a real Lagrangian density by
Wick rotating $\phi\to i\tilde\phi$, and $\tilde\phi$ will have a wrong sign
kinetic term. The Liouville field $\phi$ can be thought of as the conformal mode of gravity in two dimensions,
with metric
\begin{equation}
ds^2 = e^{2b\phi} \delta_{ab}d\sigma^a d\sigma^b.
\end{equation}
The Liouville action is generated when a two-dimensional ``matter" conformal field theory of nonzero central charge $-c$ is coupled
to gravity \cite{Polyakov:1981rd}. For the purpose of this paper, we can ignore the matter CFT, although
we shall keep in mind that the full theory of quantum gravity should have total central charge zero.

The ZZ boundary condition is such that the expectation value of $e^{2b\phi}$
goes to $+\infty$ at the boundary. Such consistent quantum boundary conditions are labeled
by a pair of positive integers $(m,n)$. There is a symmetry which exchanges $m$ with $n$ while sending $b\to 1/b$.
Global $AdS_2$ can be described as a classical solution of Liouville theory on a strip $\sigma\in(0,\pi)$, $\tau\in {\bf R}$,
with $e^{2b\phi}\to +\infty$ on the two boundaries. In the quantum theory, we can choose independently
$(m,n)$ boundary condition on the left side of the strip, and $(m',n')$ on the right side of the strip.
The Hilbert space of states on the strip will be denoted by ${\cal H}_{(m,n;m',n')}$.
It consists of boundary primary states $\psi_{k,l}$ and their Virasoro descendants. The boundary primary $\psi_{k,l}$ is characterized by
its conformal dimension
\begin{equation}
\Delta_{k,l} = \frac{Q^2}{4} - \frac{(kb+l/b)^2}{4},
\label{conf-dim}
\end{equation}
and is subject to the selection rule
\begin{equation}
\begin{aligned}
&k=|m-m'|+1,~|m-m'|+3,~\cdots~,m+m'-1; \\
&l=|n-n'|+1,~|n-n'|+3,~\cdots~,n+n'-1.
\label{sel-rules}
\end{aligned}
\end{equation}
The bulk one-point function $\langle V_\alpha(z,\bar z)\rangle$ on the disc with boundary condition
$(m,n)$, as well as the bulk-boundary two-point function (for special boundary operators), have been solved in \cite{Zamolodchikov:2001ah}.
We will need more: the bulk-boundary three-point function, boundary three-point function, and
the bulk-to-boundary four-point function. These will be solved in the next few subsections by
conformal bootstrap method.

\subsection{Fragmented $AdS_2$ as classical solutions}
It is well known that the Liouville equation of motion on the strip (for simplicity we set henceforth $\mu=1$ in the action (\ref{Liouville-action}))
\begin{equation}
(\partial^2_{\sigma}-\partial^2_t)\phi = 4\pi b e^{2 b \phi}
\label{Liouville-EOM}
\end{equation}
admits the basic static solution
\begin{equation}
\phi = -\frac{1}{2b} \ln (4\pi b^2 \sin^2 \sigma)\,.
\label{AdS2}
\end{equation}
Of course, the corresponding physical metric $ds^2=e^{2b\phi}(-dt^2+d\sigma^2)$ is nothing but the $AdS_2$ space-time. This is the $SL(2,\mathbb{R})$ invariant vacuum of Liouville theory first pointed out in \cite{D'Hoker:1982er},\cite{D'Hoker:1983is} (see also \cite{Strominger:1998yg}).

It is easy to see that the $AdS_2$ solution is part of a more general family of static solutions
\begin{equation}
\phi = -\frac{1}{2b} \ln \left(4\pi b^2 \frac{\sin^2(l\sigma)}{l^2}\right)
\label{frag-AdS2}
\end{equation}
parameterized by an integer $l \ge 1$. These solutions behave like $AdS_2$ at the $\sigma=0,\pi$ boundaries, but the metric also blows up in the ``interior" at $\sigma=\frac{p}{l}\pi\,,p=1,\cdots,l-1$. In other words, the corresponding space-time looks like $l$ disconnected copies of the $AdS_2$ solution. An example with $l=2$ is plotted in Fig.~\ref{two-AdS}. We will refer to these solutions as ``fragmented $AdS_2$ spaces".

A first hint to the relation between fragmented $AdS_2$'s and ZZ boundary primaries comes from looking at the classical Liouville stress tensor evaluated
on the solutions (\ref{frag-AdS2}). The $T_{00}$ component of the stress tensor
for a static solution reads
\begin{equation}
T_{00} = \frac{1}{4\pi} (\partial_{\sigma}\phi)^2+e^{2b\phi}-\frac{1}{2\pi b} \partial^2_{\sigma}\phi\,,
\end{equation}
where the last term comes from the ``linear dilaton" coupling to the 2d scalar curvature. Evaluated on (\ref{frag-AdS2}), this just gives the constant $T_{00}=-\frac{l^2}{4\pi b^2}$. Then one would obtain an energy relative to the $AdS_2$ vacuum
\begin{equation}
E = -\frac{l^2-1}{4b^2}\,.
\end{equation}
Note that this result precisely matches the classical limit $b\rightarrow 0$ of the conformal dimension $\Delta_{k,l}$ of the ZZ boundary primaries, eq. (\ref{conf-dim}) ($k$ drops out of the classical limit, as long as it is much smaller than $\frac{1}{b^2}$).

\subsection{The classical limit of bulk-boundary three point functions}
A given ZZ boundary primary $|\psi\rangle$ should correspond to a deformation of the Liouville profile (i.e. the space-time metric) in the bulk. Specifically, we would like to argue that the relevant bulk metrics in the classical limit $b\rightarrow 0$ correspond to the ``fragmented" $AdS_2$ spaces (\ref{frag-AdS2}). To test this idea, we shall study the expectation value $\langle \phi\rangle$ of the Liouville field on the strip, in a boundary primary state $|\psi\rangle$. This can be done by using as a ``probe" the bulk primary operator $V_{\alpha} = e^{2 \alpha \phi}$. More precisely, we need to compute the disc bulk-boundary three point function
\begin{equation}
\langle \psi(y_1) \psi(y_2) V_\alpha(z,\bar z) \rangle =
|z-\bar z|^{-2\Delta_\alpha} (y_1-y_2)^{-2h} {\cal F}(\eta)
\label{bulk-boundary-3pt}
\end{equation}
where $\Delta_\alpha=\alpha(Q-\alpha)$ is the dimension of $V_{\alpha}$, $h$ is the dimension of $\psi$, and $\eta$ is the $SL(2,{\mathbb R})$ invariant cross ratio
\begin{equation}
\eta = \frac{(z-\bar z)(y_1-y_2)}{(z-y_2)(y_1-\bar z)} = 1-e^{2i\sigma}\,,
\end{equation}
where $\sigma$ is the spatial coordinate on the strip (to obtain this relation, one can use $SL(2,\mathbb{R})$ to set $y_1=0\,,y_2=\infty$). The three point function (\ref{bulk-boundary-3pt}) is interpreted as the expectation value $\langle\psi|V_{\alpha}(\sigma)|\psi\!\rangle$ in the ZZ boundary primary $|\psi\rangle$. When $\psi$ is the identity operator, this is just the bulk one-point function computed by ZZ \cite{Zamolodchikov:2001ah}
\begin{equation}
\langle V_{\alpha}(z,\bar z) \rangle = \frac{U(\alpha)}{|z-\bar z|^{2\Delta_{\alpha}}}\,.
\end{equation}
Transforming back to strip coordinates $z=e^{i\sigma+\tau}$, one can see that in fact this is just the $AdS_2$ metric (\ref{AdS2}).

The correlation function (\ref{bulk-boundary-3pt}) depends of course on the explicit choice of $(m,n)$ boundary conditions. For now we keep the analysis general and do not specify the type of boundary conditions. Let us consider the simplest nontrivial example, $\psi= \psi_{1,2}$. According to $(\ref{conf-dim})$, it has conformal dimension
\begin{equation}
h_{1,2} = -\frac{1}{2}-\frac{3}{4 b^2}.
\end{equation}
All ZZ boundary primaries are degenerate, i.e. their conformal families contains null states. In particular, the conformal family of $\psi_{1,2}$ has a null state at level two, namely $(L_{-1}^2+
b^{-2}L_{-2})|\psi_{1,2}\rangle=0$. It follows that the bulk-boundary three point function satisfies the differential equation
\begin{equation}
\begin{aligned}
&\left\{ {\partial_{y_1}^2} + b^{-2} \left[{h_{1,2}\over (y_2-y_1)^2}
+{\Delta_\alpha\over(z-y_1)^2} + {\Delta_\alpha\over (\bar z-y_1)^2}\right. \right.
\\ &~~~~\left.\left.
- {\partial_{y_2}\over y_2-y_1} - {\partial_z\over z-y_1}-{\partial_{\bar z}\over \bar z-y_1}
 \right] \right\}\langle \psi_{1,2}(y_1) \psi_{1,2}(y_2) V_\alpha(z) \rangle=0\,.
\end{aligned}
\end{equation}
In terms of ${\cal F}(\eta)$, the equation is
\begin{equation}\label{hypo}
\eta(\eta-1) {\cal F}''(\eta) + \left[(2+b^{-2})\eta-2(1+b^{-2})\right] {\cal F}'(\eta)
+b^{-2}\Delta_\alpha {\eta\over \eta-1} {\cal F}(\eta)=0\,.
\end{equation}

\begin{figure}
\begin{center}
\includegraphics[width=100mm]{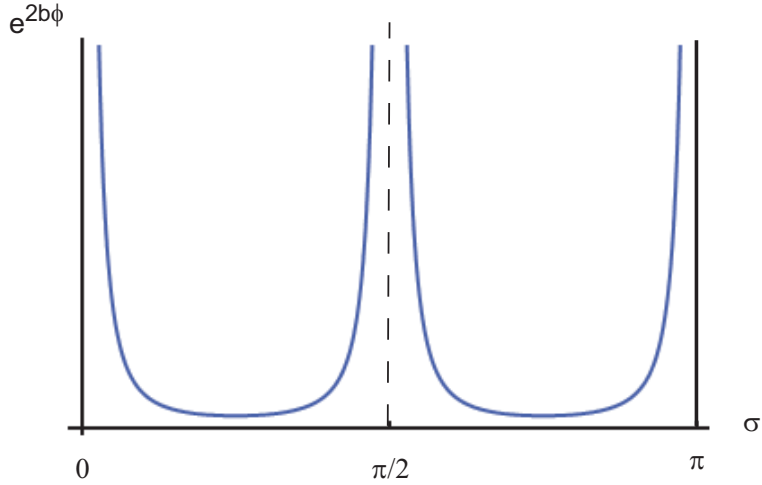}
\parbox{13cm}{
\caption{
The ``two-fragmented" $AdS_2$ space.
\label{two-AdS}}}
\end{center}
\end{figure}

In the next subsection we will explicitly solve this equation at finite $b$ and discuss in detail the results. Here we first present an easy way to arrive at the classical limit of the
bulk-boundary three point function, hence the classical interpretation of the boundary primary
$\psi_{k,l}$. The idea is that the equation (\ref{hypo}) has a naive classical limit ($b\to 0$),
\begin{equation}
(\eta-2) {\cal F}_{cl}'(\eta) + \Delta_\alpha {\eta\over\eta-1} {\cal F}_{cl}(\eta)=0\,.
\end{equation}
The solution is readily obtained
\begin{equation}
{\cal F}_{cl}(\eta=1-e^{2i\sigma}) = (\cos\sigma)^{-2\Delta_\alpha}\,,
\end{equation}
where $\Delta_\alpha \simeq \alpha/b$. Combining with the prefactor $|z-\bar z|^{2\Delta_{\alpha}}$ and transforming to the strip, one obtains as expected the two-fragmented $AdS_2$ (see Fig.~\ref{two-AdS}), i.e. $\langle\psi_{1,2}|V_{\alpha}(\sigma)|\psi_{1,2}\!\rangle \sim \left(\sin 2\sigma\right)^{-2\alpha/b}$.
Note also that since the differential equation reduces to first order, the choice of boundary
condition will not matter in this limit.

Let us now examine the bulk-boundary three point function involving $\psi_{1,3}$
\begin{equation}
\langle \psi_{1,3}(y_1) \psi_{1,3}(y_2) V_\alpha(z) \rangle =
|z-\bar z|^{-2\Delta_\alpha} (y_1-y_2)^{-2h_{1,3}} {\cal F}_{1,3}(\eta)
\end{equation}
with $h_{1,3}=-1-2b^{-2}$. The conformal family of $\psi_{1,3}$ has a null state at level 3,
\begin{equation}
\left(L_{-1}^3 + 4b^{-2} L_{-2} L_{-1} + (2b^{-2}+4b^{-4})L_{-3}\right)|\psi_{1,3}\rangle = 0
\label{null-onethree}
\end{equation}
The differential equation on the disc three point function is
\begin{equation}
\begin{aligned}
&\left\{ {\partial_{y_1}^3} + 4b^{-2} \left[{h_{1,3}\over (y_2-y_1)^2}
+{\Delta_\alpha\over(z-y_1)^2} + {\Delta_\alpha\over (\bar z-y_1)^2}
- {\partial_{y_2}\over y_2-y_1} - {\partial_z\over z-y_1}-{\partial_{\bar z}\over \bar z-y_1}
 \right]\partial_{y_1}  \right.
\\ &\left.+(2b^{-2}+4b^{-4})
\left[{2h_{1,3}\over (y_2-y_1)^3}
+{2\Delta_\alpha\over(z-y_1)^3} + {2\Delta_\alpha\over (\bar z-y_1)^3}
- {\partial_{y_2}\over (y_2-y_1)^2} - {\partial_z\over (z-y_1)^2}\right.\right.
\\
&~~~~~~~~~~\left.\left.-{\partial_{\bar z}\over (\bar z-y_1)^2}
 \right] \right\}\langle \psi_{1,3}(y_1) \psi_{1,3}(y_2) V_\alpha(z) \rangle=0\,.
\end{aligned}
\end{equation}
In the $b\to 0$ limit (with $\Delta_\alpha$ held fixed), the equation reduces to
\begin{equation}\label{tto}
(\eta-1)(\eta^2-3\eta+3) {{\cal F}_{1,3}^{cl}}'(\eta) +2\Delta_\alpha \eta(\eta-2) {\cal F}_{1,3}^{cl}(\eta)=0\,.
\end{equation}
The solution is
\begin{equation}
{\cal F}_{1,3}^{cl}(\eta=1-e^{2i\sigma}) = (1+2\cos (2\sigma))^{-2\Delta_\alpha}\,.
\end{equation}
which precisely gives rise to the 3-fragmented $AdS_2$ after the $z$-dependent prefactor is included. In fact, we observe more generally that
the conformal family of $\psi_{1,l}$ has a null state at level $l$, of the form (see page 245 of \cite{DiFrancesco:1997nk})
\be
\det\left[ - J_- + \sum_{m=0}^{l-1} b^{-2m}J_+^m L_{-m-1} \right]|\psi_{1,l}\rangle=0,
\ee
where the determinant is taken over an $l\times l$ matrix, with $J_\pm$ defined by
\be
J_- =\left. \begin{pmatrix} 0 & 0 & \cdots & 0 & 0 \\ 1 & 0 & \cdots & 0 & 0 \\
0 & 1 & \cdots & 0 & 0\\
& & \cdots &  & \\ 0 & 0&  \cdots & 1 & 0 \end{pmatrix}\right._{l\times l},~~~~~
J_+ = \left. \begin{pmatrix} 0 & l-1 & 0 & 0 & \cdots & 0 & 0 \\ 0 & 0 & 2(l-2) & 0 & \cdots & 0 & 0\\
0 & 0 & 0 & 3(l-3) & \cdots & 0 & 0\\
& & & \cdots &  & \\
0 & 0 & 0 & 0 & \cdots & 0 & l-1\\
0 & 0 & 0 & 0 & \cdots & 0 & 0 \end{pmatrix}\right._{l\times l},
\ee
In particular, the classical limit of the null state equation for $\psi_{1,l}$ is given by $L_{-l}|\psi_{1,l}\rangle={\cal O}(b^2)$. Writing
\begin{equation}
\langle \psi_{1,l}(y_1) \psi_{1,l}(y_2) V_\alpha(z) \rangle =
|z-\bar z|^{-2\Delta_\alpha} (y_1-y_2)^{-2h_{1,l}} {\cal F}_{1,l}(\eta)
\end{equation}
Analogously to (\ref{tto}), the classical constraining equation can be obtained as the first order differential equation
\begin{equation}
\eta(\eta-1){{\cal F}_{1,l}^{cl}}'(\eta) + \Delta_\alpha \left[2-\eta+l\eta {(1-\eta)^l+1\over (1-\eta)^l-1} \right]
{\cal F}_{1,l}^{cl}(\eta)=0\,.
\end{equation}
The solution is
\begin{equation}
{\cal F}_{1,l}^{cl}(\eta=1-e^{2i\sigma}) = \left({\sin l\sigma\over\sin \sigma}\right)^{-2\Delta_\alpha}
\end{equation}
Consequently,
\begin{equation}
\langle \psi_{1,l}| e^{2\alpha\phi(\sigma)} |\psi_{1,l}\rangle \sim (\sin l\sigma)^{-2\alpha/b},
\end{equation}
corresponding to the $l$-fragmented $AdS_2$, in accordance with our general proposal.

\subsection{Quantum bulk-boundary three-point functions}
\subsubsection{General boundary condition}

We shall now study the quantum bulk-boundary three point function (\ref{bulk-boundary-3pt}) at finite coupling. We will specialize to the simplest non-trivial example $\psi=\psi_{1,2}$ (in section \ref{degenerate-probes} we will propose a method to obtain the bulk-boundary three point function for general $\psi_{k,l}$). To this purpose, we need to solve the second order differential equation (\ref{hypo}) exactly at finite $b$. The equation can be put in the standard hypergeometric form, and the general solution is
\begin{equation}
\label{solhyp}
\begin{aligned}
{\cal F}(\eta) &= c_1 (1-\eta)^{\alpha/b} {}_2F_1({2\alpha\over b},1+b^{-2};2+2b^{-2};\eta)
\\
&+ c_2 (1-\eta)^{\alpha/b} \eta^{-1-2b^{-2}} {}_2F_1(-1-2b^{-2}+{2\alpha\over b},-b^{-2};-2b^{-2};\eta)\,,
\end{aligned}
\end{equation}
where ${}_2F_1(A,B;C;z)$ is the Gauss hypergeometric function.
The constants $c_1$ and $c_2$ are related to the factorization of the disc three point function
along the boundary operator channels corresponding to the boundary primaries $\mathbf{1}$ (the identity operator) and $\psi_{1,3}$ (if it is allowed by the specific choice of boundary condition, according to the selection rules (\ref{sel-rules})).
In particular,
\begin{equation}\begin{aligned}
&c_1= U(\alpha), \\
&c_2= \langle \psi_{1,2}\psi_{1,2}\psi_{1,3} \rangle R_{1,3}(\alpha),
\end{aligned}\end{equation}
where $U(\alpha)$ is the coefficient of the disc one point function of $V_\alpha$,
and $R_{1,3}(\alpha)$ is the coefficient of the bulk-to-boundary two point function of $V_\alpha$ with $\psi_{1,3}$.
By $\langle \psi_{1,2}\psi_{1,2}\psi_{1,3}\rangle$ we mean the coefficient of the corresponding
boundary three point function, with the appropriate boundary conditions along the three segments of the
boundary of the disc in between the operator insertions. Note that $\psi_{1,3}$ has conformal dimension $h_{1,3}=-1-2b^2$, $R_{1,3}(0)=0$, and that both $R_{1,3}(\alpha)$ and $\langle \psi_{1,2}\psi_{1,2}\psi_{1,3}\rangle$ depend on the boundary conditions. The explicit expressions for $U(\alpha)$ and $R_{1,3}(\alpha)R_{1,3}(-b/2)$ were derived by ZZ \cite{Zamolodchikov:2001ah}. The boundary three point function $\langle \psi_{1,2}\psi_{1,2}\psi_{1,3}\rangle$, however, was not previously derived and will be obtained below.

Let us consider the classical/weak coupling limit of (\ref{solhyp}), i.e. small $b$. The asymptotics of the Gauss hypergeometric functions can be extracted using the quadratic transformation
\begin{equation}
{}_2F_1(A,B;2B;z) = (1-z)^{-A/2} {}_2F_1({A\over2}, B-{A\over2}; B+{1\over 2}; {z^2\over 4(z-1)})
\end{equation}
and the asymptotic expansion \cite{Temme}
\begin{equation}
{}_2F_1(A,B+\lambda; C+\lambda; z) = (1-z)^{-A} (1+{\cal O}(\lambda^{-1})),~~~~\lambda\to\infty.
\end{equation}
We then find in the small $b$ limit
\begin{equation}
\begin{aligned}
\!\!\!\!\!\!\!\!
{\cal F}(\eta) &\sim U(\alpha) (1-{\eta^2\over 4(\eta-1)})^{-\alpha/b}
+\langle \psi_{1,2}\psi_{1,2}\psi_{1,3} \rangle R_{1,3}(\alpha) (1-\eta)^{{1\over 2}+b^{-2}} \eta^{-1-2b^{-2}} (1-{\eta^2\over 4(\eta-1)})^{\alpha/b-{1\over 2}}
\\ &=U(\alpha) (\cos\sigma)^{-2\alpha/b}+\langle \psi_{1,2}\psi_{1,2}\psi_{1,3} \rangle R_{1,3}(\alpha) (-2i\sin\sigma)^{-1-2b^{-2}} (\cos\sigma)^{2\alpha/b-1}\,.
\end{aligned}\end{equation}
In particular, in the $\alpha\to 0$ limit, ${\cal F}(\eta)\to 1$ as expected.
After a conformal transformation back to the strip, we conclude that
\begin{equation}\begin{aligned}\label{ress}
\langle \psi_{1,2} | e^{2\alpha \phi(\sigma)} |\psi_{1,2}\rangle &\sim
U(\alpha) (\sin 2\sigma)^{-2\alpha/b}\\
&+\langle \psi_{1,2}\psi_{1,2}\psi_{1,3} \rangle R_{1,3}(\alpha) (-i)^{-1-2b^{-2}}
(2\sin\sigma)^{-1-2b^{-2}-{2\alpha/b}} (\cos\sigma)^{2\alpha/b-1}
\end{aligned}\end{equation}
as $b\to 0$ with $\alpha/b$ fixed. This can be compared to the vacuum expectation value of $V_\alpha$, which as discussed in the previous section corresponds to the regular $AdS_2$ profile
\begin{equation}
\langle 1 | e^{2\alpha \phi(\sigma)} |1\rangle \sim U(\alpha)(2\sin \sigma)^{-2\alpha/b}\,.
\end{equation}
If the second term in (\ref{ress}) is absent, i.e. ignoring the contribution from the $\psi_{1,3}$
channel, then the contribution from the identity operator channel
suggests indeed that the boundary primary $\psi_{1,2}$ creates a state that would
correspond classically to two copies of global $AdS_2$ glued together, as predicted by the ``naive" classical limit of the differential equation discussed in the previous section. The $\psi_{1,3}$
contribution is sensitive to the boundary conditions, and will be analyzed in the next subsection for a specific choice of boundary type.

%For real small $b$, however, the $\psi_{1,3}$ channel in (\ref{ress}) will dominate
%and ruin this interpretation. This is related to the fact that $\psi_{1,3}$
%has negative dimension $-1-2b^2$. On the other hand, if we analytically continue $b=i\beta$
%and Wick rotate the Liouville field $\phi$, then the $\psi_{1,3}$ channel contribution
%may be suppressed in the classical limit (small real $\beta$), and we would then conclude that
%$\psi_{1,2}$ creates a 2-fragmented $AdS_2$. This will be confirmed below.

\subsubsection{$(1,1;1,2)$ boundary condition}

\begin{figure}
\vskip -3cm
\begin{center}
\includegraphics[width=50mm]{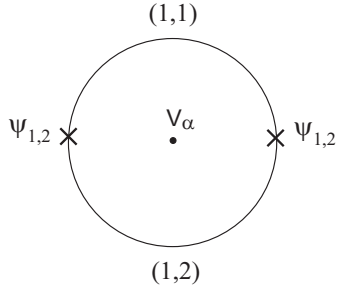}
\vskip -0.8cm
\parbox{13cm}{
\caption{
Depiction of the bulk-boundary three point function $\langle \psi_{1,2}(y_1) \psi_{1,2}(y_2) V_\alpha(z) \rangle$ with a specific choice of boundary condition.
\label{2ptfunction}}}
\end{center}
\end{figure}

\begin{figure}
\begin{center}
\includegraphics[width=70mm]{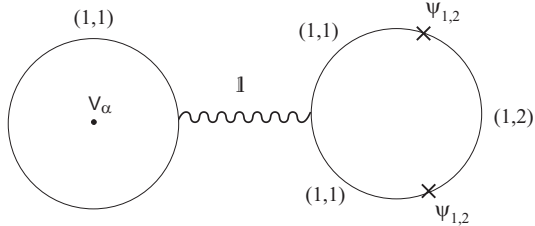}
\parbox{13cm}{
\caption{
Factorization of the bulk-boundary three point function along the $(1,1;1,1)$ channel.
\label{2ptfactor1}}}
\end{center}
\end{figure}

Let us now specialize to the strip with $(1,1)$ boundary condition on the left
and $(1,2)$ boundary condition on the right. The only allowed boundary primary operator/state is $\psi_{1,2}$.
To compute the expectation value of the Liouville field in this state, we need to compute the
bulk-boundary three point function $\langle \psi_{1,2}(y_1) \psi_{1,2}(y_2) V_\alpha(z) \rangle$,
with $(1,1)$ boundary condition on one segment of the boundary circle and $(1,2)$ boundary on the other
segment of the circle, between $y_1$ and $y_2$,  as shown in Fig.~\ref{2ptfunction}.
There are two different ways to factorize $\langle \psi_{1,2}(y_1) \psi_{1,2}(y_2) V_\alpha(z) \rangle$ into the product of a boundary three point function and a bulk-boundary two point function, along channels of $(1,1;1,1)$ boundary type or $(1,2;1,2)$ boundary type.
In the first factorization, as shown in Fig.~\ref{2ptfactor1}, the only boundary primary operator in the channel is the identity operator. We have then
\begin{equation}\label{expra}
\langle \psi(y_1) \psi(y_2) V_\alpha(z) \rangle =
|z-\bar z|^{-2\Delta_\alpha} (y_1-y_2)^{-2h} U_{1,1}(\alpha) (1-\eta)^{\alpha/b} {}_2F_1({2\alpha\over b},1+b^{-2};2+2b^{-2};\eta)\,.
\end{equation}
In the second factorization, we have the identity operator as well as $\psi_{1,3}$ propagating through the channel, as depicted in Fig.~\ref{2ptfactor1}, giving
\begin{equation}\label{exprb}
\begin{aligned}
\langle \psi(y_1) \psi(y_2) V_\alpha(z) \rangle =
|z-\bar z|^{-2\Delta_\alpha} (y_1-y_2)^{-2h} \left[U_{1,2}(\alpha) (1-\bar\eta)^{\alpha/b} {}_2F_1({2\alpha\over b},1+b^{-2};2+2b^{-2};
\bar\eta)\right.\\
\left. -ie^{-i\pi/b^2}\langle \psi_{1,2}\psi_{1,2}\psi_{1,3}\rangle R_{1,3}(\alpha) (1-\bar\eta)^{\alpha/b}
\bar\eta^{-1-2b^{-2}} {}_2F_1(-1-2b^{-2}+{2\alpha\over b},-b^{-2};-2b^{-2};\bar\eta)\right],
\end{aligned}
\end{equation}
where $\bar\eta$ is the complex conjugate of $\eta$.
One may also replace $\bar\eta$ by $\eta/(\eta-1)$, and use the property of Gauss hypergeometric functions to rewrite (\ref{exprb})
in terms of the same functions with argument $\eta$. The phase factor
$-ie^{-i\pi/b^2}$ in the second term on the RHS is such that in the factorization limit $\eta\to i\epsilon$
($\sigma\to \pi-\epsilon$),
the conformal block corresponding to the $\psi_{1,3}$ is real and positive.
\begin{figure}
\begin{center}
\includegraphics[width=70mm]{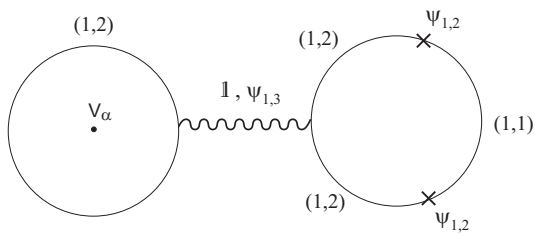}
\parbox{13cm}{
\caption{
Factorization of the bulk-boundary three point function along the $(1,2;1,2)$ channel.
\label{2ptfactor2}}}
\end{center}
\end{figure}
One may seem to run into a puzzle here, since the two ways of factorizing the bulk-boundary three point function should give the same result. The resolution is that in fact (\ref{expra}) and (\ref{exprb}) are related by analytic continuation across the branch cut of the hypergeometric
function from $\eta=1$ to infinity. This can be shown
using the monodromy of the hypergeometric function around $\eta=1$, or equivalently
\begin{equation}
\begin{aligned}
&{}_2 F_1(a,b;c;x+i\epsilon) = e^{2\pi i(a+b-c)} {}_2 F_1(a,b;c;x) \\
&+ 2\pi ie^{\pi i(a+b-c)} {\Gamma(c)\over\Gamma(a+b+1-c) \Gamma(c-a)
\Gamma(c-b)}\; {}_2 F_1(a,b;a+b+1-c;1-x)
\end{aligned}
\label{branch-cut}
\end{equation}
for real $x>1$, together with the boundary three point function $\langle \psi_{1,2}\psi_{1,2}\psi_{1,3}\rangle$ which will be explicitly computed below.

\begin{figure}
\begin{center}
\includegraphics[width=50mm]{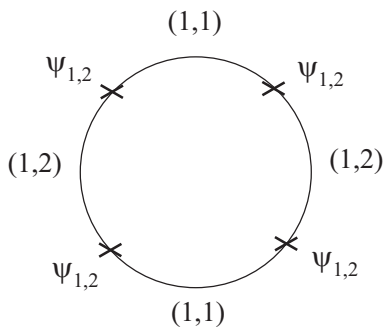}
\parbox{13cm}{
\caption{
Boundary four point function with alternating boundary conditions.
\label{4ptfunction}}}
\end{center}
\end{figure}

\begin{figure}
\begin{center}
\includegraphics[width=100mm]{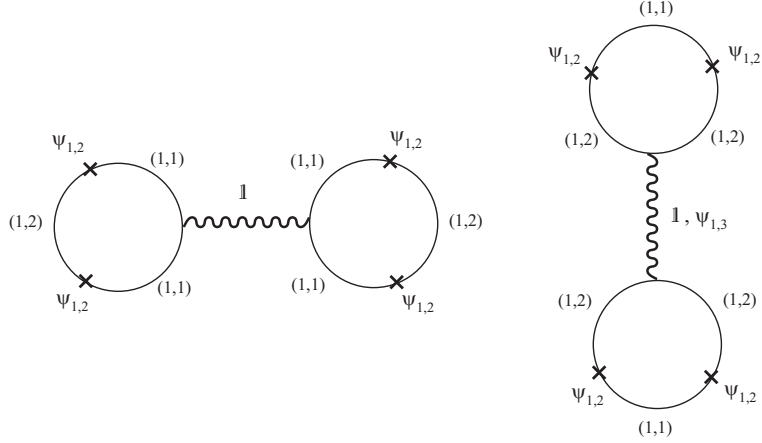}
\vskip -1.5cm
\parbox{13cm}{
\caption{
The two possible factorizations of the boundary four point function.
\label{4ptfactor}}}
\end{center}
\end{figure}

To compute $\langle \psi_{1,2}\psi_{1,2}\psi_{1,3} \rangle$, we make use of
the boundary four point function \break
$\langle \psi_{1,2}(y_1) \psi_{1,2}(y_2) \psi_{1,2}(y_3) \psi_{1,2}(y_4) \rangle$, with alternating $(1,1)$ and $(1,2)$ boundary conditions
along the four segments of the boundary circle separated by the boundary operators, see Fig.~\ref{4ptfunction}.
It is determined by a function ${\cal G}(\zeta)$,
\begin{equation}
\begin{aligned}
\langle \psi_{1,2}(y_1) \psi_{1,2}(y_2) \psi_{1,2}(y_3) \psi_{1,2}(y_4) \rangle &=
(y_1-y_2)^{-2h_{1,2}} (y_3-y_4)^{-2h_{1,2}} {\cal G}(\zeta),\\
\zeta &= {(y_1-y_2)(y_3-y_4)\over (y_3-y_2)(y_1-y_4)}\,.
\end{aligned}
\end{equation}
${\cal G}(\zeta)$ satisfies the same differential equation as that of ${\cal F}(\eta)$,
with $\Delta_\alpha$ replaced by $h$ ($\alpha\to-{1\over 2b}$), $\eta$ replaced by $\zeta$.
The solutions takes the form
\begin{equation}\label{solhypa}
\begin{aligned}
{\cal G}(\zeta) &= c_1 (1-\zeta)^{-{1\over 2}b^{-2}} {}_2F_1(-b^{-2},1+b^{-2};2+2b^{-2};\zeta)
\\
&+ c_2 (1-\zeta)^{-{1\over 2}b^{-2}}
\zeta^{-1-2b^{-2}} {}_2F_1(-1-3b^{-2},-b^{-2};-2b^{-2};\zeta)\,.
\end{aligned}
\end{equation}
%Channel duality implies
%\begin{equation}\label{channel}
%\left( {\zeta\over 1-\zeta} \right)^{-2h} {\cal G}(\zeta) = {\cal G}(1-\zeta).
%\end{equation}
Again, we can factorize it into two boundary three point functions, along either $(1,1;1,1)$ channel (with the only primary being the
identity operator) or $(1,2;1,2)$ channel (with primaries ${\bf 1}$ and $\psi_{1,3}$). The two factorizations are shown in Fig.~\ref{4ptfactor}.
The first factorization gives
\begin{equation}\label{expraa}
\begin{aligned}
\langle \psi_{1,2}(y_1) \psi_{1,2}(y_2) \psi_{1,2}(y_3) \psi_{1,2}(y_4) \rangle &=
(y_1-y_2)^{-2h_{1,2}} (y_3-y_4)^{-2h_{1,2}} \\
\times& (1-\zeta)^{-{1\over 2}b^{-2}} {}_2F_1(-b^{-2},1+b^{-2};2+2b^{-2};\zeta)\,,
\end{aligned}
\end{equation}
while the second factorization gives
\begin{equation}\label{exprbb}
\begin{aligned}
&\langle \psi_{1,2}(y_1) \psi_{1,2}(y_2) \psi_{1,2}(y_3) \psi_{1,2}(y_4) \rangle =
C(b) (y_1-y_2)^{-2h_{1,2}} (y_3-y_4)^{-2h_{1,2}} \left( \frac{1-\zeta}{\zeta} \right)^{-2h_{1,2}} \\
&~~~\times \left[\zeta^{-{1\over 2}b^{-2}} {}_2F_1(-b^{-2},1+b^{-2};2+2b^{-2};1-\zeta)
\right.\\ &\left.~~~+ \langle \psi_{1,2}\psi_{1,2}\psi_{1,3} \rangle^2\, \zeta^{-{1\over 2}b^{-2}}
(1-\zeta)^{-1-2b^{-2}} {}_2F_1(-1-3b^{-2},-b^{-2};-2b^{-2};1-\zeta)\right]
\end{aligned}
\end{equation}
where $C(b)=-(2 \cos(\pi b^{-2}))^{-1}$ is a normalization factor (which can be determined by matching the two channels as explained below). This nontrivial normalization factor is due to the different boundary conditions on the channels of the two factorizations. In fact, we can identify
\begin{equation}
C(b) = {C_{1,1}(b)\over C_{1,2}(b)}
\end{equation}
where $C_{m,n}(b)$ stands for the disc amplitude with no insertions and $(m,n)$ boundary condition.
The forms of (\ref{expraa}) and (\ref{exprbb}) agree by the identity
\begin{equation}\label{identities}
\begin{aligned}
{}_2F_1(A,B;C;z) &= {\Gamma(C)\Gamma(C-A-B)\over \Gamma(C-A)\Gamma(C-B)}{}_2F_1(A,B;A+B+1-C;1-z)
\\&+ {\Gamma(C)\Gamma(A+B-C)\over \Gamma(A)\Gamma(B)} (1-z)^{C-A-B} {}_2F_1(C-A,C-B;1+C-A-B;1-z) \\
& = (1-z)^{C-A-B}{}_2F_1(C-A,C-B;C;z).
\end{aligned}
\end{equation}
Using this, we then derive the boundary three point function
\begin{equation}
\langle \psi_{1,2}\psi_{1,2}\psi_{1,3} \rangle = \pm \left[-2\cos({\pi\over b^2})
{\Gamma(1+{2\over b^2}) \Gamma(2+{2\over b^2})\over \Gamma(1+{1\over b^2}) \Gamma(2+{3\over b^2})} \right]^{1\over 2}.
\label{boundary-3pt}
\end{equation}
In order to match (\ref{expra}) and (\ref{exprb}) through analytic continuation, as explained above, we need to choose the negative sign in (\ref{boundary-3pt}).
Using the results of \cite{Zamolodchikov:2001ah}, it follows that
\begin{equation}
{\langle \psi_{1,2}\psi_{1,2}\psi_{1,3} \rangle R_{1,3}(\alpha)\over U_{1,1}(\alpha)}
= -{8\over\pi}\left(1+{2\over b^2}\right) \sin(2\pi {\alpha\over b}) \sin (2\pi {\alpha-b^{-1}\over b})
{\Gamma({2\over b^2})^2 \Gamma(1-{2\alpha\over b})\Gamma(-1-{2\over b^2}+{2\alpha\over b})\over
\Gamma({1\over b^2})^2 }\,.
\label{R}
\end{equation}
It is also useful to note the identity
\begin{equation}
{U_{1,2}(\alpha)\over U_{1,1}(\alpha)} = {\cos(\pi({2\alpha\over b} - {1\over b^2}))\over \cos(\pi/b^2)}\,.
\label{UU}
\end{equation}
Using (\ref{R}) and (\ref{UU}), remarkably, one can check that (\ref{exprb}) is indeed related to (\ref{expra}) by analytic continuation to a different sheet across its branch cut. This also provides a check of the result of \cite{Zamolodchikov:2001ah} for $U_{m,n}(\alpha)$ and $R_{1,3}(\alpha)$.

The quantum bulk-boundary three-point function can therefore be determined by analytically continuing
(\ref{expra}) from $\sigma=0$ to $\sigma=\pi$. In practice, such analytic continuation may be defined by ``patching" (\ref{expra}) to (\ref{exprb}) at $\sigma = \pi/2$, while using the standard definition of the hypergeometric functions with their conventional branch cuts. On the two halves of the strip, we find in the $b\to 0$ limit, with $\alpha/b$ finite,
\begin{equation}
\begin{aligned}
& \langle \psi_{1,2} | e^{2\alpha\phi(\sigma)}|\psi_{1,2}\rangle \sim U_{1,1}(\alpha) (\sin 2\sigma)^{-2\alpha/b},
~~~~\sigma\in (0,{\pi\over 2}),\\
&\langle \psi_{1,2} | e^{2\alpha\phi(\sigma)}|\psi_{1,2}\rangle \sim U_{1,2}(\alpha) (\sin 2\sigma)^{-2\alpha/b}
\\
&~~~~~+ \langle \psi_{1,2}\psi_{1,2}\psi_{1,3}\rangle R_{1,3}(\alpha)(2\sin\sigma)^{-1-2b^{-2}-2\alpha/b}
(\cos\sigma)^{2\alpha/b-1},~~~~\sigma\in ({\pi\over 2},\pi).
\end{aligned}
\end{equation}
We see that for real $b$ and generic values of $\alpha$, in the classical limit
the $\psi_{1,3}$ channel dominates for $\sigma>\pi/2$, and appears to
``erase" the right $AdS_2$. The exceptional cases are when the probe bulk operator has
$\alpha=-nb/2$ for a positive integer $n$, and $R_{1,3}(\alpha)$ vanishes. In this case the hypergeometric
function reduces to elementary functions. For instance, when $\alpha=-b/2$, we have
\begin{equation}
\langle \psi_{1,2} | e^{-b\phi(\sigma)}|\psi_{1,2}\rangle = {\sin 2\sigma\over 2} (\sin\sigma)^{{3\over 2}b^2}
\end{equation}
agreeing with the ``naive" classical limit of two-fragmented $AdS_2$.

\begin{figure}
\begin{center}
\includegraphics[width=140mm]{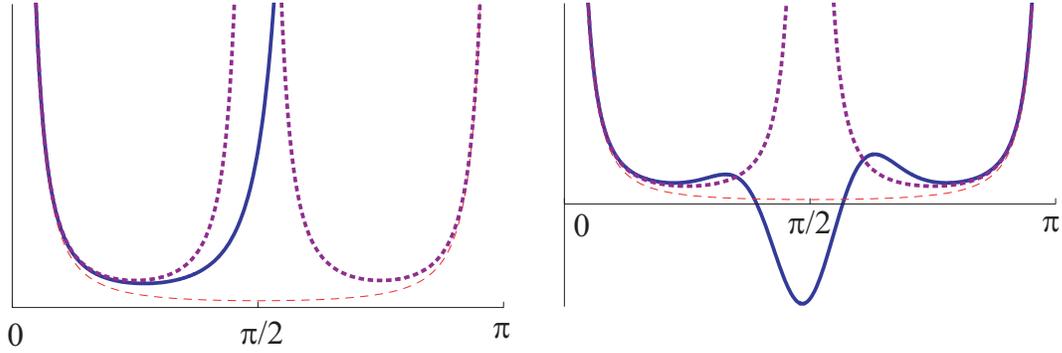}
\parbox{13cm}{
\caption{Plots of the bulk-boundary 3-point function $\langle\psi_{1,2}|V_{\alpha}(\sigma)|\psi_{1,2}\!\rangle$ (with $\alpha=b$) at finite coupling, for $b$ real (left) and $b$ imaginary (right), with $|b|=0.3$. The dashed line represents the $AdS_2$ metric, while the dotted one corresponds to the two-fragmented $AdS_2$. With real $b$ (left) the asymptotic $AdS_2$ boundary condition is respected only at the $\sigma=0$ boundary, and the classical limit $b\rightarrow 0$ produces a single $AdS_2$ fragment (the solid line extends to $\sigma=\pi$ and erases the second $AdS_2$). On the other hand, with imaginary $b$ (right) the profile is asymptotically $AdS_2$ at both boundaries and the limit $b\rightarrow 0$ gives the two-fragmented $AdS_2$ metric.
\label{3pt-plots}}}
\end{center}
\end{figure}

On the other hand, we can consider $b=i\beta$ purely imaginary, and take $\alpha$ to be purely imaginary as well (or equivalently, Wick rotating the Liouville field
$\phi$). In the classical limit $\beta\to 0$
($\alpha/b$ taken to be real and finite),
the identity channel dominates the $\psi_{1,3}$ channel, and we have
\begin{equation}
\begin{aligned}
& \langle \psi_{1,2} | e^{2\alpha\phi(\sigma)}|\psi_{1,2}\rangle \sim U_{1,1}(\alpha) (\sin 2\sigma)^{-2\alpha/b},
~~~~\sigma\in (0,{\pi\over 2}),\\
&\langle \psi_{1,2} | e^{2\alpha\phi(\sigma)}|\psi_{1,2}\rangle \sim U_{1,2}(\alpha) (\sin 2\sigma)^{-2\alpha/b},
~~~~\sigma\in ({\pi\over 2},\pi),
\end{aligned}
\end{equation}
i.e. the expectation value of $e^{2\alpha\phi(\sigma)}$ scales like $(\sin2\sigma)^{-2\alpha/b}$ on both halves of the strip, leading to two-fragmented $AdS_2$. Plots of the analytically continued bulk-boundary three point function for real and purely imaginary $b$ are given in Fig.~\ref{3pt-plots}.

We see that near the mid point $\sigma=\pi/2$ where the two fragmented $AdS_2$'s meet, quantum correction is
large despite that the conformal dimension of the probe operator $\Delta_\alpha\sim \alpha/b\ll |c|$. In particular, the expectation value
of $V_\alpha(\sigma=\pi/2)$ in the state $|\psi_{1,2}\rangle$ is given by (using the quadratic transform of
${}_2F_1$)
\begin{equation}
\begin{aligned}
\langle \psi_{1,2}|e^{2\alpha\phi(\pi/2)}|\psi_{1,2}\rangle &= {}_2F_1({\alpha\over b},
1+b^{-2}-{\alpha\over b};{3\over 2}+b^{-2};1) \\ &= {\sqrt{\pi}\,\Gamma({3\over 2}+b^{-2})\over \Gamma({1\over 2}+{\alpha\over b})
\Gamma({3\over 2}+b^{-2}-{\alpha\over b})}\\
&\to \left\{ \begin{array}{ll}  b^{-2\alpha/b}{\sqrt{\pi}\over \Gamma({1\over 2}+{\alpha\over b})},~~~~&b\to +0, \\
(-b^2)^{-\alpha/b} {\cos(\pi({\alpha\over b}-b^{-2}))\over
\cos(\pi b^{-2})} {\sqrt{\pi}\over \Gamma({1\over 2}+{\alpha\over b})},~~~~&b\to i 0. \end{array} \right.
\end{aligned}
\end{equation}
For instance, for imaginary $b$, $\langle \psi_{1,2}|e^{2b\phi(\sigma)}|\psi_{1,2}\rangle$ is negative at $\sigma=\pi/2$, as
in figure \ref{3pt-plots}.

\begin{figure}
\begin{center}
\includegraphics[width=140mm]{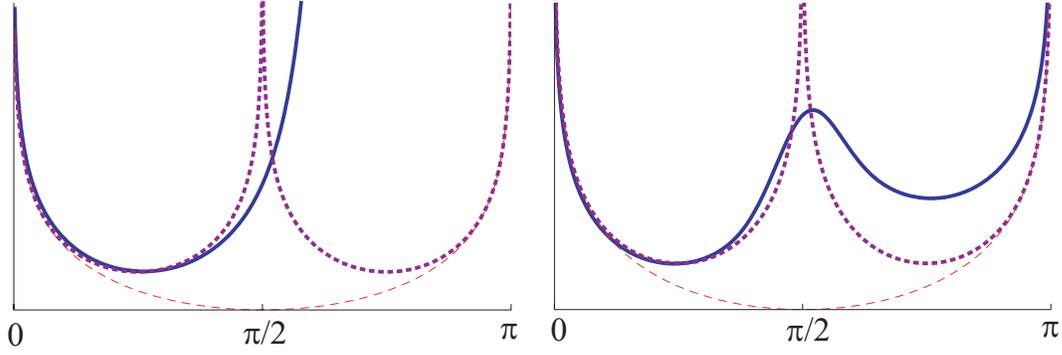}
\parbox{13cm}{
\caption{Plots of the expectation value of $\phi(\sigma)$ in the ZZ boundary primary $|\psi_{1,2}\rangle$, for $b$ real (left) and $b$ imaginary (right), with a generic non-integer value of
$|b^{-2}|$. The dashed line represents the Liouville profile of
the $AdS_2$ vacuum, while the dotted one corresponds to the two-fragmented $AdS_2$.
In the case of imaginary $b$, the profile of the Liouville field on the right $AdS_2$ is shifted by the constant
${2\pi\over b}\tan(\pi/b^2)$.
\label{vevphi-b-no-int}}}
\end{center}
\end{figure}

It is natural to consider $V_\alpha$ in the $\Delta_\alpha\sim \alpha/b\to 0$ limit. We find for $b$ purely imaginary,
in the $b\to i0$ limit,
\begin{equation}
\begin{aligned}
\langle \psi_{1,2}|\phi(\sigma)|\psi_{1,2}\rangle &= {1\over 2}\left.{\partial\over \partial{\alpha}}
\langle \psi_{1,2}|V_\alpha(\sigma)|\psi_{1,2}\rangle\right|_{\alpha=0} \\
&\to \left\{\begin{array}{ll} -{1\over b} \ln|\sin 2\sigma|+const, ~~~~&\sigma\in (0,{\pi\over 2}), \\
-{1\over b} \ln|\sin 2\sigma|+{2\pi\over b}\tan(\pi/b^2)+const,~~~~&\sigma\in ({\pi\over 2},\pi), \end{array}\right.
\end{aligned}
\end{equation}
where the overall constant shift can be absorbed into the Liouville cosmological constant. Curiously,
the profile of the Liouville field in the two $AdS_2$'s differ by a constant shift ${2\pi\over b}\tan(\pi/b^2)$, coming from
the derivative of $U_{1,2}(\alpha)/U_{1,1}(\alpha)$ at $\alpha=0$,
which is oscillatory as $b\to 0$. At the special values $b={i/\sqrt{n}}$, for positive integer $n$, this shift
is absent and we have a regular semi-classical limit as $n\to \infty$. This suggests a quantization of the central charge in
Liouville $AdS_2$ gravity, $c=1+6Q^2=13-6(n+{1\over n})$.

\begin{figure}
\begin{center}
\includegraphics[width=140mm]{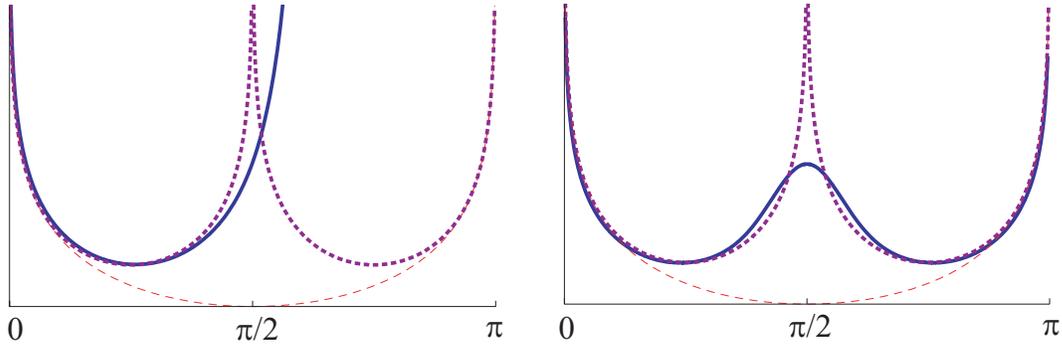}
\parbox{13cm}{
\caption{Plots of the expectation value of $\phi(\sigma)$ in the ZZ boundary primary $|\psi_{1,2}\rangle$, for $b$ real (left) and $b$ imaginary (right), with the integer value of
$|b^{-2}|=16$. The dashed line represents the Liouville profile of
the $AdS_2$ vacuum, while the dotted one corresponds to the two-fragmented $AdS_2$.
\label{vevphi-b-int}}}
\end{center}
\end{figure}

To summarize the results of this section, we found that:

\noindent (1) For real values of $b$, in the $b\to 0$ limit, only one of the two $AdS_2$ fragments survives in the quantum theory,
and the geometry of the ZZ boundary primary $\psi_{1,2}$ is asymptotically $AdS_2$ only near the $(1,1)$ boundary, while
destroying the $AdS_2$ boundary condition on the $(1,2)$ boundary. However, the bulk operators
$V_{-nb/2}$ for positive integer $n$ still see the two-fragmented $AdS_2$.

\noindent (2) For purely imaginary values of $b$, the ZZ boundary primary $\psi_{1,2}$
creates two-fragmented $AdS_2$ in the semi-classical limit, which survives
in the quantum theory. A regular semi-classical limit also suggests the quantization of the
Liouville central charge, $b=i/\sqrt{n}$ and $c=1+6Q^2=13-6(n+{1\over n})$, where $n$ is a positive integer.

Perhaps it is worth recalling here that purely imaginary $b$ is the required choice if one wishes to consistently couple the Liouville sector to a unitary matter CFT, in the semi-classical limit.

\subsection{Probing $|\psi_{m,n}\rangle$ with degenerate bulk primaries}
\label{degenerate-probes}

In this subsection, we consider the bulk-boundary three point function involving the first
degenerate bulk primary $V_{-b/2}$ and general ZZ boundary primary $\psi_{m,n}$ with $(1,1)$
boundary condition on one side of the disc and $(m,n)$ boundary condition on the other side,
\begin{equation}
\langle \psi_{m,n}(y_1) \psi_{m,n}(y_2) V_{-b/2}(z,\bar z) \rangle =
|z-\bar z|^{1+{3\over 2}b^2} y_{12}^{-2h_{m,n}} {\cal F}(\eta)\,.
\end{equation}
The null state at level 2 in the conformal family of $V_{-b/2}$ gives rise to a second order
differential equation on ${\cal F}(\eta)$. The two independent solutions to the
differential equation are conformal blocks corresponding to the factorization on
the identity operator and $\psi_{1,3}$. Since we have chosen the $(1,1)$ boundary condition
at $\sigma=0$, then the factorization through $\psi_{1,3}$ channel is absent
as $V_{-b/2}$ approaches the left boundary. This fixes the solution to
\begin{equation}
{\cal F}(\eta) = (1-\eta)^{n+1+(m+1)b^2\over 2} {}_2F_1(n+1+(m+1)b^2,1+b^2;
2+2b^2;\eta)
\end{equation}
or in terms of the expectation value of $V_{-b/2}$ on the strip,
\begin{equation}
\begin{aligned}
\langle \psi_{m,n}| e^{-b\phi(\sigma)} | \psi_{m,n}\rangle
&= (\sin\sigma)^{1+{3\over 2}b^2} e^{i(n+1+(m+1)b^2)\sigma}
{}_2F_1(n+1+(m+1)b^2,1+b^2;2+2b^2;1-e^{2i\sigma})
\\
&\to {\sin (n\sigma)\over n}~~~~(b\to 0)\,,
\end{aligned}
\end{equation}
confirming the interpretation of $\psi_{m,n}$ as $n$-fragmented $AdS_2$ in the
semi-classical limit. Although, we should note that we expect the same subtlety in the case of real $b$ discussed in the previous section, where generic $V_\alpha$ will only see one of the $n$ $AdS_2$'s, the other fragments being ``erased" by quantum effects. For purely imaginary $b$, however, we expect the $n$-fragmented $AdS_2$ to survive in the full quantum theory.

\section{Interactions of fragmented $AdS_2$}

\subsection{Boundary three-point functions}
\label{gen-3pt}

Let us denote by $\langle m,n,k\rangle$ the boundary three-point function
$\langle \psi_{1,m} \psi_{1,n} \psi_{1,k}\rangle$ with boundary condition of $(1,2)$ type between
$\psi_{1,m}$ and $\psi_{1,n}$ insertions, $(1,n-1)$ between $\psi_{1,n}$ and $\psi_{1,k}$, and
$(1,m-1)$ between $\psi_{1,m}$ and $\psi_{1,k}$. Note that $\langle m,n,k\rangle$
is not symmetric in $n,k,m$. In the classical limit, however, we have seen that the profile of
$\psi_{1,m}$ is not sensitive to the boundary types, provided that the primary $\psi_{1,m}$ is contained in the
Hilbert space of the given boundary types. So we expect that
the classical limit of $\langle m,n,k\rangle$ to be symmetric in $m,n,k$. We will find that this is indeed the case, apart from
an oscillating factor. Also note that $\langle m,n,k\rangle$ is nonzero only when $|m-n|+1\leq k\leq
m+n-3$ and $m+n+k+1\in 2{\mathbb Z}$, due to the selection rule (\ref{sel-rules}).

We shall consider the boundary four-point function $\langle 2,m,k,n\rangle$, with boundary condition
$(1,1; 1,m; 1,n-1;1,2)$ around the boundary circle. It can factorize as
\begin{equation}
\langle \psi_{1,2}\psi_{1,m}\psi_{1,k}\psi_{1,n}\rangle \to
\langle \psi_{1,2}\psi_{1,n}\psi_{1,n-1}\rangle \langle \psi_{1,n-1}\psi_{1,2}
\psi_{1,m}\rangle
\end{equation}
or as (schematically)
\begin{equation}
\langle \psi_{1,2}\psi_{1,m}\psi_{1,k}\psi_{1,n}\rangle \to
\langle \psi_{1,2}\psi_{1,m}\psi_{1,m-1}\rangle \langle \psi_{1,m-1}\psi_{1,n}
\psi_{1,k}\rangle
+\langle \psi_{1,2}\psi_{1,m}\psi_{1,m+1}\rangle \langle \psi_{1,m+1}\psi_{1,n}
\psi_{1,k}\rangle\,.
\end{equation}
Writing
\begin{equation}
\langle \psi_{1,2}(y_1)\psi_{1,m}(y_2)\psi_{1,k}(y_4)\psi_{1,n}(y_3)\rangle =
(y_{12}y_{34})^{\sum h_i} \left(\prod_{1\leq i<j\leq 4} y_{ij}^{-h_i-h_j}\right)
{\cal F}(\eta),
\end{equation}
where $\eta=y_{12}y_{34}/y_{14}y_{32}$, ${\cal F}(\eta)$ obeys the hypergeometric equation
coming from the null state in the conformal family of $\psi_{1,2}$. The general solution is
\begin{equation}
\begin{aligned}
{\cal F}(\eta) = (1-\eta)^{(1-{n\over 2})(1+{n\over 2b^2})}\eta^{-1+{n+k\over 2}+{n^2+k^2-2m\over 4b^2}}
\left[C_1\; {}_2F_1({n-m-k\over 2b^2},1+{n+k-m\over 2b^2};-{m\over b^2};\eta)\right.\\
\left. +C_2\eta^{1+{m\over b^2}}\; {}_2F_1(1+{n+m-k\over 2b^2},2+{n+m+k\over 2b^2};2+{m\over b^2};\eta)\right].
\end{aligned}
\end{equation}
The limit $\eta\to 0$ corresponds to the factorization through $\psi_{m\pm 1}$, whereas $\eta\to 1$ corresponds to
the factorization though $\psi_{n-1}$. Imposing that there is no factorization through $\psi_{n+1}$
(as required by our choice of boundary condition), we find
\begin{equation}
\begin{aligned}
&{\langle m+1,2,m\rangle \langle m+1,n,k\rangle\over \langle m-1,2,m\rangle \langle m-1,n,k\rangle}
={C_1\over C_2} = {\Gamma(2+{m\over b^2}) \Gamma({k-m-n\over 2b^2})\Gamma(-1-{k+m+n\over 2b^2})
\over \Gamma(-{m\over b^2}) \Gamma(1+{k+m-n\over 2b^2}) \Gamma({m-n-k\over 2b^2})} \\
&={\cos((n+k-3m){\pi\over 2b^2})-\cos((n+k+m){\pi\over 2b^2})\over\cos
 ({k\pi\over b^2}) -\cos({(m+n)\pi\over b^2})}
 {\Gamma(1+{m\over b^2}) \Gamma(2+{m\over b^2}) \Gamma(1+{n+k-m\over 2b^2})
\over\Gamma(1+{k+m-n\over 2b^2}) \Gamma(1+{m+n-k\over 2b^2})\Gamma(2+{k+m+n\over 2b^2})}.
\end{aligned}
\end{equation}
Choosing $n=2,k=m$, we obtain $\langle m+1,2,m\rangle / \langle m-1,2,m\rangle$; then we can further derive
$\langle m+1,n,k\rangle / \langle m-1,n,k\rangle$. We shall not write the general formula, but
focus on the classical limit ($b\to 0$),
\begin{equation}
\begin{aligned}
&{\langle m+1,n,k\rangle \over \langle m-1,n,k\rangle} \sim (oscillating~factor)\times
\exp\left[{1\over b^2} (m \log(4m)+{1\over 2}(m-1)\log(m-1) \right.\\ &\left.+{1\over2}(m+1)\log(m+1)- {1\over 2}(m+k-n) \log(m+k-n)
- {1\over 2}(m-k+n) \log(m-k+n)\right.\\ &\left.
+ {1\over 2}(-m+k+n) \log(-m+k+n)
- {1\over 2}(m+k+n) \log(m+k+n)) +{\cal O}(1)\right].
\end{aligned}
\end{equation}
Iterating this relation, we find
\begin{equation}
\begin{aligned}
&{\langle 1+x,1+y,1+z\rangle \over \langle -1+x,-1+y,-1+z\rangle} \sim (oscillating~factor)\times
\exp\left[{1\over b^2} (x \log(4x)+{1\over 2}(x-1)\log(x-1)
\right.\\ &\left.
+{1\over 2}(x+1)\log(x+1) +y \log(4y)+{1\over 2}(y-1)\log(y-1)+{1\over 2}(y+1)\log(y+1)
\right.\\ &\left.
+z \log(4z)+{1\over 2}(z-1)\log(z-1)+{1\over 2}(z+1)\log(z+1)
\right.\\ &\left.- {1\over 2}(x+y-z) \log(x+y-z)
- {1\over 2}(x-y+z) \log(x-y+z)\right.\\ &\left.
- {1\over 2}(-x+y+z) \log(-x+y+z)
- {1\over 2}(x+y+z) \log(x+y+z)\right.\\ &\left.
- {1\over 2}(x+y+z-2) \log(x+y+z-2)- {1\over 2}(x+y+z+2) \log(x+y+z+2)) +{\cal O}(1)\right].
\end{aligned}
\label{analytic-cont}
\end{equation}
This expression is particularly interesting because, as we will show in section \ref{geod-approx} below, an analytic continuation to non-integer $x,y,z$ can be matched against the geodesic approximation of three point particles in $AdS_2$.
%Analytically continuing to non-integer $x,y,z$, in the limit of small $x,y,z$, this will be matched to the instanton %action of
%the geodesics approximation of 3 particles of masses $-x/b^2, -y/b^2, -z/b^2$. Note that
%that conformal dimension of say $\psi_{1,1+x}$ is $\Delta_{1,1+x}= -{1\over 4b^2}((1+x)^2-1)+{\cal O}(1)
%\simeq -{x\over 2b^2}$. WHY THE FACTOR OF 2 DIFFERENCE??

We can also give a closed form expression in the limit $x,y,z\gg 1$, corresponding to the scattering
of $AdS_2$'s with many fragments. In this case, we get
\begin{equation}
\begin{aligned}
&\langle x,y,z \rangle \sim \exp \left[{1\over b^2} (x^2\log x+y^2\log y+z^2\log z -{(x+y-z)^2\over 4}\log(x+y-z)\right.\\
&\left.
-{(x-y+z)^2\over 4}\log(x-y+z)-{(-x+y+z)^2\over 4}\log(-x+y+z)-{(x+y+z)^2\over 4}\log(x+y+z)) \right].
\end{aligned}
\end{equation}

\subsection{Bulk-boundary four-point functions}
\label{3pt-instantons}

\begin{figure}
\begin{center}
\includegraphics[width=50mm]{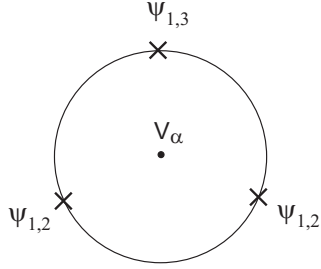}
\vskip -1cm
\parbox{13cm}{
\caption{
The bulk-boundary four point function.
\label{3ptfunction}}}
\end{center}
\end{figure}

In this section we study the disc bulk-boundary four-point function $\langle \psi_{1,2}(y_1)
\psi_{1,2}(y_2)\psi_{1,3}(y_3) V_\alpha(z,\bar z)\rangle$, see Fig.~\ref{3ptfunction}. The choice of boundary type is not important for now, since we will be interested in the classical limit of this correlation function.
By conformal invariance, this four-point function takes the form
\begin{equation}
\langle \psi_{1,2}(y_1)
\psi_{1,2}(y_2)\psi_{1,3}(y_3) V_\alpha(z,\bar z)\rangle =
|z-\bar z|^{-2\Delta_\alpha} y_{12}^{-2h_{1,2}+h_{1,3}} (y_{13}y_{23})^{-h_{1,3}} {\cal F}(\eta,\bar\eta)
\end{equation}
where
\begin{equation}
\eta = {(z-y_1)y_{23}\over (z-y_3)y_{21}},~~~~\bar\eta = {(\bar z-y_1)y_{23}\over (\bar z-y_3)y_{21}}.
\end{equation}
We can use the $SL(2,\mathbb{R})$ symmetry to fix for example $y_1=0,y_2=1$ and $y_3=\infty$. Then $\eta$ and $\bar\eta$ simply coincide with the coordinates $z$ and $\bar z$ parameterizing the position of the ``probe" bulk primary $V_{\alpha}(z,\bar z)$.

\begin{figure}
\begin{center}
\includegraphics[width=50mm]{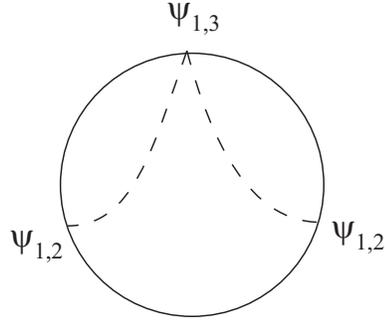}
\parbox{13cm}{
\caption{
Schematic depiction of the classical solution (\ref{inst-223}), corresponding to three Poincar\'{e} discs patched together along the dashed lines.
\label{3ptfrag}}}
\end{center}
\end{figure}

\begin{figure}
\begin{center}
\includegraphics[width=60mm]{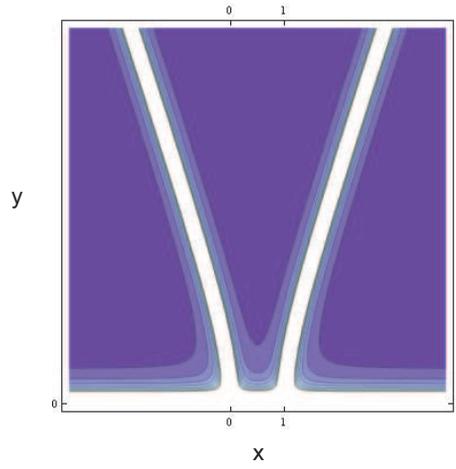}
\parbox{13cm}{
\caption{
Contour plot of the solution (\ref{inst-223}) in the upper half plane coordinates. The points $x=0$ and $x=1$ on the real axis correspond to the insertions of the two $\psi_{1,2}$ operators. The ``fragmentation lines'' are described by eq. (\ref{fraglines}).
\label{223contour}}}
\end{center}
\end{figure}

\begin{figure}
\begin{center}
\includegraphics[width=70mm]{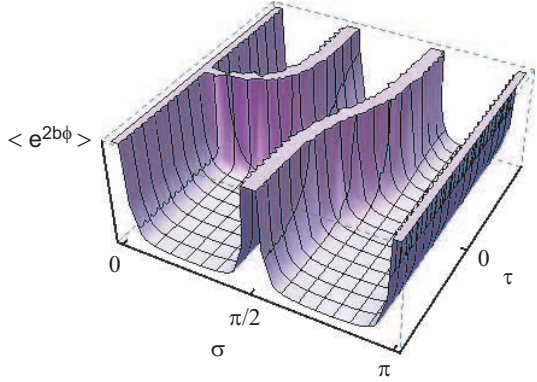}
\parbox{13cm}{
\caption{
Strip coordinates plot of the classical Liouville profile corresponding to the boundary three point function $<\psi_{1,2}\psi_{1,2}\psi_{1,3}>$. The strip $\tau \in \mathbb{R}$,$\,0 \le \sigma \le \pi$ fragments into three disconnected pieces.
\label{223strip}}}
\end{center}
\end{figure}

The constraining equation from the level 2 null state in the conformal family of $\psi_{1,2}(y_1)$ reduces to
a second order differential equation on ${\cal F}$,
\begin{equation}
\begin{aligned}
&\Delta_\alpha (\eta-\bar\eta)^2 {\cal F} + \bar\eta^2 \eta(\eta-1) (2(1+b^2)\eta+1) \partial_\eta {\cal F}
+ \eta^2 \bar\eta(\bar\eta-1) (2(1+b^2)\bar\eta+1) \partial_{\bar\eta} {\cal F} \\
&+b^2 \eta^2\bar\eta^2 \left[(\eta-1)^2 \partial_{\eta}^2{\cal F}+ (\bar\eta-1)^2 \partial_{\bar\eta}^2{\cal F}
+ 2|\eta-1|^2 \partial_\eta\partial_{\bar\eta} {\cal F} \right]=0\,.
\end{aligned}
\end{equation}
In the classical limit $b \rightarrow 0$ (with $\alpha \over b$ fixed), this reduces to the first order equation
\begin{equation}
\Delta_\alpha (\eta-\bar\eta)^2 {\cal F}^{cl} + \bar\eta^2 \eta(\eta-1) (2\eta+1) \partial_\eta {\cal F}^{cl}
+ \eta^2 \bar\eta(\bar\eta-1) (2\bar\eta+1) \partial_{\bar\eta} {\cal F}^{cl} =0\,.
\end{equation}
Similarly, there is another equation coming from $\psi_{1,2}(y_2)$, which is identical to the above equation with
${\cal F}(\eta,\bar\eta)$ replaced by ${\cal F}(1-\eta,1-\bar\eta)$. The solution to this pair of equations is
(up to a normalization constant)
\begin{equation}
{\cal F}^{cl}(\eta,\bar\eta) = \left[|\eta|^2|1-\eta|^2 (2\eta^2+2\bar\eta^2+2|\eta|^2-3\eta-3\bar\eta)^{-2} \right]^{\Delta_\alpha}.
\end{equation}
This means that if the classical limit of the three-point function $\langle \psi_{1,2}(y_1)
\psi_{1,2}(y_2)\psi_{1,3}(y_3)\rangle$ is dominated by an instanton solution, the solution has Liouville profile
\begin{equation}
\langle e^{2\alpha\phi}\rangle_{inst} = |z-\bar z|^{-2\alpha/b}
\left[|z|^2|1-z|^2 (2z^2+2\bar z^2+2|z|^2-3z-3\bar z)^{-2} \right]^{\alpha/b}\,.
\end{equation}
The instanton solution has ``physical" metric (we fix the overall normalization to agree with the conventions of ZZ)
\begin{equation}
e^{2b\phi} d z d\bar z = {36|z|^2|1-z|^2 dz d\bar z\over
\pi b^2|z-\bar z|^2 (2z^2+2\bar z^2+2|z|^2-3z-3\bar z)^{2}}.
\label{inst-223}
\end{equation}
This is indeed a solution to Liouville equation, and corresponds to three Poincar\'{e} discs patched together, depicted schematically in Fig.~\ref{3ptfrag}.
In the upper half plane coordinate $z=x+i y$, the three disconnected $AdS_2$'s are glued along the two curves
\begin{equation}
y = \sqrt{3x(x-1)},~~~~x<0 ~{\rm or}~x>1.
\label{fraglines}
\end{equation}
A contour plot of the classical solution (\ref{inst-223}) in the upper half plane coordinates, showing the curves (\ref{fraglines}) is shown in Fig.~\ref{223contour}.
It is also interesting to visualize the solution in the strip coordinates defined by $z=e^{i\sigma+\tau}$. The corresponding plot is shown in Fig.~\ref{223strip}.

It is actually not difficult to obtain the classical instanton profile for more general boundary three point functions. Consider for example the four point function $\langle \psi_{1,3}(y_1)
\psi_{1,3}(y_2)\psi_{1,3}(y_3) V_\alpha(z,\bar z)\rangle$. Again, by conformal invariance we can write
\begin{equation}
\langle \psi_{1,3}(y_1)
\psi_{1,3}(y_2)\psi_{1,3}(y_3) V_\alpha(z,\bar z)\rangle =
|z-\bar z|^{-2\Delta_\alpha} (y_{12}y_{13}y_{23})^{-h_{1,3}} {\cal F}_{333}(\eta,\bar\eta)
\end{equation}
where $\eta$ and $\bar\eta$ are defined as above. The constraining equation from the null state in the conformal family of $\psi_{1,3}(y_1)$, see eq. (\ref{null-onethree}), reduces in the classical limit to the first order differential equation
\begin{equation}
2\Delta_\alpha (\eta-\bar\eta)^2 (|\eta|^2-\eta-\bar\eta){\cal F}^{cl}_{333} + \eta \bar\eta^3 (\eta^2-1) (2\eta-1) \partial_\eta {\cal F}^{cl}_{333}
+ \bar\eta \eta^3 (\bar\eta^2-1) (2\bar\eta-1) \partial_{\bar\eta} {\cal F}^{cl}_{333} =0\,,
\end{equation}
and as before there is a similar equation coming from $\psi_{1,3}(y_2)$. The solution to this couple of first order differential equations (up to an overall constant) turns out to be
\begin{equation}
{\cal F}^{cl}_{333}(\eta,\bar\eta) = \left[\frac{|\eta|^2|1-\eta|^2}{\left( \eta^3+\bar\eta^3-2(\eta^2+\bar\eta^2)(|\eta|^2+1)+|\eta|^2(5\eta+5\bar\eta
-2|\eta|^2-2)\right)^2} \right]^{\Delta_\alpha}\,,
\end{equation}
and the physical instanton metric ($\Delta_\alpha=1$) corresponding to the three point function $\langle \psi_{1,3}\psi_{1,3}\psi_{1,3}\rangle$ is therefore
\begin{equation}
e^{2b\phi} dz d\bar z = {36|z|^2|1-z|^2 dz d\bar z\over
\pi b^2|z-\bar z|^2 \left( z^3+\bar z^3-2(z^2+\bar z^2)(|z|^2+1)+|z|^2(5z+5\bar z-2|z|^2-2)\right)^2}.
\label{inst-333}
\end{equation}
One can verify that this is a solution to Liouville equation, and as expected corresponds to four copies of the Poincar\'{e} disc patched together.

\subsection{Instantons interpolating fragmented $AdS_2$'s}
\label{general-inst}

It is instructive to write the above solutions in the general form
\begin{equation}
e^{2b\phi} =\frac{1}{\pi b^2} \frac{\partial A(z) \bar\partial B(\bar z)}{(1-A(z)B(\bar z))^2}\,,
\label{gen-sol}
\end{equation}
which is in fact the most general solution of Liouville equation
\begin{equation}
\partial\bar \partial \phi-\pi b e^{2 b\phi}=0\,.
\end{equation}
For example, the regular $AdS_2$ solution corresponds to $B(\bar z)=1/A(\bar z)$ and $A(z)=z$, while the $n$-fragmented solution has $A(z)=z^n$. In fact, this is perhaps the fastest way to see that the stress tensor for the $n$-fragmented $AdS_2$ agrees with the conformal dimension of the operator $\psi_{1,n}$ (one simply looks at the Schwartzian derivative of the conformal transformation $w=z^n$).

Going back to the ``instanton" solutions (\ref{inst-223}),(\ref{inst-333}) found above, one can see that they take indeed the form (\ref{gen-sol}) (still with $B(\bar z)=1/A(\bar z)$, which ensures that the metric is real). By direct calculation one finds that $\partial A(z)=z(z-1)$ for the $\langle\psi_{1,2}\psi_{1,2}\psi_{1,3}\rangle$ and $\partial A(z)=\frac{z^2(z-1)^2}{(z-\frac{1}{2})^2}$ for the $\langle\psi_{1,3}\psi_{1,3}\psi_{1,3}\rangle$ case. Motivated by this, we conjecture that the general instanton solution corresponding to the three point function $\langle\psi_{1,n}(0)\psi_{1,m}(1)\psi_{1,k}(\infty)\rangle$ is given by \footnote{The metric (\ref{gen-sol}) with $B=1/\bar A$ is invariant under the $SL(2,\mathbb{R})$ transformation $A \rightarrow (a A+b)/(c A+d)$, $\partial A \rightarrow \partial A /(c A + d)^2$. One can reach the standard form (\ref{Aprime}) by applying such $SL(2,\mathbb{R})$ transformation.}
\begin{equation}
\partial A(z) = \frac{z^{n-1}(z-1)^{m-1}}{P(z)^2}\,,
\label{Aprime}
\end{equation}
where $P(z)$ is a $(n+m-k-1)/2$-degree polynomial with distinct roots (to be determined below), namely $P(z) = \prod_{i=1}^{(n+m-k-1)/2}(z-z_i)$. Note that $(n+m-k-1)/2$ is an integer as implied by the three point function selection rules. The conjecture is motivated as follows. First, the degree of the zeroes at $z=0$ and $z=1$ and of the pole at $z=\infty$ are fixed by demanding that near those points the metric looks respectively like the $n$-fragmented, the $m$-fragmented and the $k$-fragemented $AdS_2$. Furthermore, the fact that $P(z)$ has distinct roots and the denominator of (\ref{Aprime}) is $P(z)^2$  follows by requiring that near each of the $z_i$ the metric looks like the regular $AdS_2$. To see this, consider for simplicity $\partial A(z) \sim \frac{1}{(z-z_0)^s}$ near $z=z_0$. Then writing $z=z_0 + r e^{i\theta}$, the metric close to $z_0$ takes the form $$ds^2 \sim \frac{(dr/r)^2 + d\theta^2}{\sin^2((s-1)\theta)}\,,$$ so that we have to choose $s=2$ as claimed. Finally, whe have to specify the position of the roots $z_i$ of $P(z)$. If we insist that $A(z)$ has to be a rational function, which seems to be a natural assumption, then the roots can be determined by requiring that the poles at $z=z_i$ have vanishing residue, namely
\begin{equation}
\frac{d}{dz}\left[(z-z_i)^2 \partial A(z)\right]|_{z=z_i}=0\,~~~~~~i=1,\cdots,\frac{n+m-k-1}{2}\,.
\label{roots}
\end{equation}
It is easy to verify that the above solutions for $\langle\psi_{1,2}\psi_{1,2}\psi_{1,3}\rangle$ and $\langle\psi_{1,3}\psi_{1,3}\psi_{1,3}\rangle$ satisfy the conjecture (\ref{Aprime}). By directly solving for the classical limit of the bulk-boundary four point function as described above, we have also successfully checked the conjecture on a few other explicit examples such as $\langle\psi_{1,2}\psi_{1,3}\psi_{1,4}\rangle$ and $\langle\psi_{1,3}\psi_{1,3}\psi_{1,5}\rangle$, which respectively have $\partial A(z)=z(z-1)^2$ and $\partial A(z) = z^2(z-1)^2$. Other tests of (\ref{Aprime}),(\ref{roots}) come from these known examples by applying an inversion $z\rightarrow 1/z$ to map the origin to infinity. For example, one finds that the solution for $\langle\psi_{1,3}(0)\psi_{1,2}(1)\psi_{1,2}(\infty)\rangle$ has $\partial A(z) = \frac{z^2(z-1)}{(z-2/3)^2}$ in agreement with the conjecture. A further check  comes from $\langle\psi_{1,5}(0)\psi_{1,3}(1)\psi_{1,3}(\infty)\rangle$, which turns out to be given by $\partial A(z) = \frac{z^4(z-1)^2}{(z-z_1)^2(z-\bar z_1)^2}$ with $z_1=\frac{1}{20}(15+i\sqrt{15})$, as predicted by $(\ref{Aprime}),(\ref{roots})$.

\subsection{Comparison with the geodesic approximation}
\label{geod-approx}

Consider the classical limit $b \rightarrow 0$ of the boundary three point function, for example eq. (\ref{boundary-3pt}). In this limit one gets
\begin{equation}\label{classical-3pt}
\begin{aligned}
&\langle \psi_{1,2}\psi_{1,2}\psi_{1,3} \rangle \sim \pm  \frac{2\sqrt 2}{3^{\frac{3}{4}}} \left(-\cos \frac{\pi}{b^2}\right)^{\frac{1}{2}} \, e^{-\frac{1}{2 b^2} \ln \frac{27}{16}},~~~~{\rm real} ~b\to 0,
\\
&\langle \psi_{1,2}\psi_{1,2}\psi_{1,3} \rangle \sim \pm  \frac{\sqrt 2}{3^{\frac{3}{4}}} \left(-{1+2\cos\frac{2\pi}{b^2}\over \cos \frac{\pi}{b^2}}\right)^{\frac{1}{2}} \, e^{-\frac{1}{2 b^2} \ln \frac{27}{16}},~~~~{\rm imaginary} ~b\to 0.
\end{aligned}
\end{equation}
The exponential term suggests that it should be possible to obtain this result by a semiclassical gravity calculation.\footnote{Curiously, for a special set of ``quantized" values of $b$, namely $b^{-2}=n$ being an odd integer,
the oscillatory factor on the RHS of (\ref{classical-3pt}) is a constant independent $n$.}
Namely one should evaluate the regularized action on the ``instanton" solution, i.e. the classical Liouville profile (\ref{inst-223}), corresponding to the insertion of the three boundary primary operators, which was obtained in section \ref{3pt-instantons}. In this section we present a different calculation based on point particles moving along geodesics in $AdS_2$, which interestingly matches a particular analytic continuation of the boundary three point function discussed in Section \ref{gen-3pt}.

\begin{figure}
\begin{center}
\includegraphics[width=100mm]{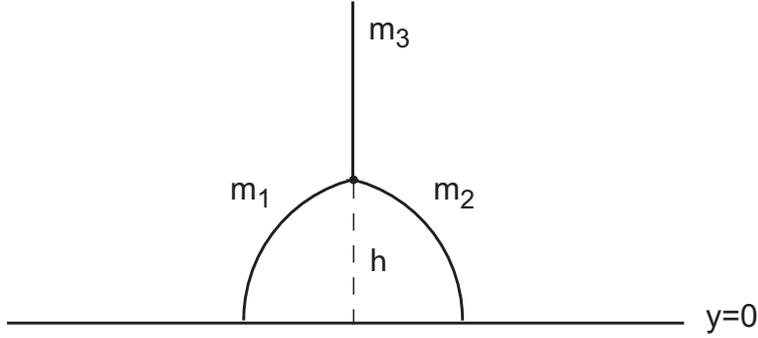}
\parbox{13cm}{
\caption{
Three point particles of masses $m_1,m_2,m_3$ moving along geodesics in the Poincar\`{e} half plane.
\label{geodesics}}}
\end{center}
\end{figure}

Consider three point particles of masses $m_1,m_2,m_3$ starting off at the boundary of the disk and moving along geodesics until they meet at one point in the interior. The geodesic approximation is expected to be valid when the mass of the particles are large compared to the $AdS$ scale, and
the gravity coupling is weak, i.e. in the limit $1 \ll m_i \ll 1/b^2$ (in $AdS$ units). Mapping the problem to the upper half plane, we can place the particles with masses $m_1$ and $m_2$ on the real line (separated say by a distance $L$), and the particle with mass $m_3$ at $i \infty$. The particles $m_1$ and $m_2$ move along circles and $m_3$ along a straight line, until the geodesics meet at a certain height $h$, as shown in Fig.~\ref{geodesics}.
Using as variables the final angular position $\theta_1$ and $\theta_2$ of the circular geodesics, the total action for the system reads
\begin{equation}
S = m_1 \int_{\frac{\epsilon}{R_1}}^{\theta_1} \frac{d\theta}{\sin\theta} +m_2 \int_{\frac{\epsilon}{R_2}}^{\theta_2} \frac{d\theta}{\sin\theta} + m_3 \ln \frac{\Lambda}{h}\,,
\end{equation}
where we have introduced cutoffs $\epsilon \rightarrow 0$ and $\Lambda \rightarrow \infty$. The geometry implies
\begin{equation}
h(\theta_1,\theta_2) = \frac{L}{\tan\frac{\theta_1}{2} + \tan\frac{\theta_2}{2}}\, ~~~~~R_{1,2} = \frac{h}{\sin\theta_{1,2}}\,.
\end{equation}
Then one gets
\begin{equation}
\begin{aligned}
S=&(m_1+m_2-m_3) \ln h -m_1 \ln \cos^2 \frac{\theta_1}{2} - m_2 \ln \cos^2 \frac{\theta_2}{2} \\
&-(m_1+m_2) \ln \epsilon + m_3 \ln \Lambda\,.
\end{aligned}
\label{part-action}
\end{equation}
It is not difficult to extremize this action with respect to $\theta_1$ and $\theta_2$ for general $m_1,m_2,m_3$. The solution turns out to be
\begin{equation}
\tan^2\frac{\theta_1}{2}=\frac{m_2^2-(m_1-m_3)^2}{(m_1+m_3)^2-m_2^2}\,,
~~~~~~~~~
\tan^2\frac{\theta_2}{2}=\frac{m_1^2-(m_2-m_3)^2}{(m_2+m_3)^2-m_1^2}\,.
\end{equation}

Plugging back into (\ref{part-action}) and removing the divergencies, one finds the following general formula for the regularized action
\begin{equation}
\begin{aligned}
&S_{reg} = {1\over b^2} \left[ \sum_{i=1}^3 m_i\log(2m_i) -{1\over 2}(m_1+m_2-m_3)\log(m_1+m_2-m_3) \right.\\ &\left.
-{1\over 2}(m_1-m_2+m_3)\log(m_1-m_2+m_3)-{1\over 2}(-m_1+m_2+m_3)\log(-m_1+m_2+m_3)\right.\\ &\left.-{1\over 2}(m_1+m_2+m_3)\log(m_1+m_2+m_3)\right]\,.
\end{aligned}
\end{equation}
This expression matches the classical limit of the boundary three point function (\ref{analytic-cont}) if we take the masses to be $m_1=-x/b^2,m_2=-y/b^2,m_3=-z/b^2$, with small real $x,y,z$. Note that the conformal dimension corresponding to a scalar of mass $m$  is $h=\frac{1}{2}+\frac{1}{2}\sqrt{1+m^2}$ (where $m$ is expressed in units of the $AdS$ radius) \cite{Strominger:1998yg}. In our limit $h \sim \frac{m}{2}$, consistently with the dimension $\Delta_{1,1+x} \sim -\frac{x}{2b^2}$ of the ``analytically continued" operator $\psi_{1,1+x}$.

\subsection*{Acknowledgments}
We are grateful to D. Gaiotto, G. Moore, A. Pakman and D. Shih for useful discussions, and especially to A. Strominger for collaborations at the initial stage of this work and on related topics. X.Y. would like to thank  Tata Insititute of Fundamental Research and the organizers of the Monsoon Workshop on
String Theory for hospitality during the completion of this work.
The work of S.G. is supported in part by the Center for the Fundamental Laws of Nature at Harvard University and by NSF grants PHY-024482 and DMS-0244464.
The work of X.Y. is supported by a Junior Fellowship from the Harvard Society of Fellows.


\begin{thebibliography}{}

%\cite{Ginsparg:1993is}
\bibitem{Ginsparg:1993is}
  P.~H.~Ginsparg and G.~W.~Moore,
  ``Lectures on 2-D gravity and 2-D string theory,''
  arXiv:hep-th/9304011.
  %%CITATION = HEP-TH/9304011;%%

%\cite{Strominger:1998yg}
\bibitem{Strominger:1998yg}
  A.~Strominger,
  ``AdS(2) quantum gravity and string theory,''
  JHEP {\bf 9901}, 007 (1999)
  [arXiv:hep-th/9809027].
  %%CITATION = JHEPA,9901,007;%%

%\cite{Maldacena:1998uz}
\bibitem{Maldacena:1998uz}
  J.~M.~Maldacena, J.~Michelson and A.~Strominger,
  ``Anti-de Sitter fragmentation,''
  JHEP {\bf 9902}, 011 (1999)
  [arXiv:hep-th/9812073].
  %%CITATION = JHEPA,9902,011;%%

%\cite{Strominger:2003tm}
\bibitem{Strominger:2003tm}
  A.~Strominger,
  ``A matrix model for AdS(2),''
  JHEP {\bf 0403}, 066 (2004)
  [arXiv:hep-th/0312194].
  %%CITATION = JHEPA,0403,066;%%
  
%\cite{Hartman:2008dq}
\bibitem{Hartman:2008dq}
  T.~Hartman and A.~Strominger,
  ``Central Charge for $AdS_2$ Quantum Gravity,''
  JHEP {\bf 0904}, 026 (2009)
  [arXiv:0803.3621 [hep-th]].
  %%CITATION = JHEPA,0904,026;%%

%\cite{Sen:2008vm}
\bibitem{Sen:2008vm}
  A.~Sen,
  ``Quantum Entropy Function from AdS(2)/CFT(1) Correspondence,''
  arXiv:0809.3304 [hep-th].
  %%CITATION = ARXIV:0809.3304;%%

%\cite{Sen:2008yk}
\bibitem{Sen:2008yk}
  A.~Sen,
  ``Entropy Function and AdS(2)/CFT(1) Correspondence,''
  JHEP {\bf 0811}, 075 (2008)
  [arXiv:0805.0095 [hep-th]].
  %%CITATION = JHEPA,0811,075;%%

%\cite{Zamolodchikov:2001ah}
\bibitem{Zamolodchikov:2001ah}
  A.~B.~Zamolodchikov and A.~B.~Zamolodchikov,
  ``Liouville field theory on a pseudosphere,''
  arXiv:hep-th/0101152.
  %%CITATION = HEP-TH/0101152;%%

%\cite{Fateev:2000ik}
\bibitem{Fateev:2000ik}
  V.~Fateev, A.~B.~Zamolodchikov and A.~B.~Zamolodchikov,
  ``Boundary Liouville field theory. I: Boundary state and boundary  two-point
  function,''
  arXiv:hep-th/0001012.
  %%CITATION = HEP-TH/0001012;%%

%\cite{Seiberg:1990eb}
\bibitem{Seiberg:1990eb}
  N.~Seiberg,
  ``Notes on quantum Liouville theory and quantum gravity,''
  Prog.\ Theor.\ Phys.\ Suppl.\  {\bf 102}, 319 (1990).
  %%CITATION = PTPSA,102,319;%%

%\cite{Nakayama:2004vk}
\bibitem{Nakayama:2004vk}
  Y.~Nakayama,
  ``Liouville field theory: A decade after the revolution,''
  Int.\ J.\ Mod.\ Phys.\  A {\bf 19}, 2771 (2004)
  [arXiv:hep-th/0402009].
  %%CITATION = IMPAE,A19,2771;%%

%\cite{Polyakov:1981rd}
\bibitem{Polyakov:1981rd}
  A.~M.~Polyakov,
  ``Quantum geometry of bosonic strings,''
  Phys.\ Lett.\  B {\bf 103}, 207 (1981).
  %%CITATION = PHLTA,B103,207;%%

%\cite{D'Hoker:1982er}
\bibitem{D'Hoker:1982er}
  E.~D'Hoker and R.~Jackiw,
  ``Liouville Field Theory,''
  Phys.\ Rev.\  D {\bf 26}, 3517 (1982).
  %%CITATION = PHRVA,D26,3517;%%

%\cite{D'Hoker:1983is}
\bibitem{D'Hoker:1983is}
E.~D'Hoker, D.~Z.~Freedman and R.~Jackiw,
``SO(2,1) Invariant Quantization Of The Liouville Theory,''
Phys.\ Rev.\  D {\bf 28}, 2583 (1983).
%%CITATION = PHRVA,D28,2583;%%

\bibitem{Temme}
N.~M.~Temme,
``Large parameter cases of the Gauss hypergeometric function,''
Journal of Computational and Applied Mathematics {\bf 153} (2003), 441-462.
%%CITATION = PHRVA,D28,2583;%%

%\cite{DiFrancesco:1997nk}
\bibitem{DiFrancesco:1997nk}
  P.~Di Francesco, P.~Mathieu and D.~Senechal,
  ``Conformal Field Theory,''
%\href{http://www.slac.stanford.edu/spires/find/hep/www?irn=3810356}{SPIRES entry}
{\it  New York, USA: Springer (1997) 890 p}


\end{thebibliography}
\end{document}